\documentclass{article}

\usepackage{arxiv}

\usepackage[utf8]{inputenc} 
\usepackage[T1]{fontenc}    
\usepackage{hyperref}       
\usepackage{url}            
\usepackage{booktabs}       
\usepackage{amsfonts}       
\usepackage{nicefrac}       
\usepackage{microtype}      
\usepackage{lipsum}
\usepackage{graphicx}

\usepackage{booktabs} 
\usepackage{amsmath,amssymb,amsfonts}
\usepackage{algorithmic}
\usepackage{textcomp}
\usepackage[inline]{enumitem}
\usepackage{enumitem}
\newlist{steps}{enumerate}{1}
\setlist[steps, 1]{label = Step \arabic*:}
\graphicspath{ {Figures/} {Experiments/} }

\usepackage{tikz}
\usepackage{pgfplots}
\usepackage{pgfplotstable}
\usepackage{subfigure}
\usetikzlibrary{matrix}
\usepgfplotslibrary{groupplots}
\pgfplotsset{compat=newest}

\newcommand{\wtg}{$\mathcal{W}2\mathcal{G}$}
\newcommand{\gtw}{$\mathcal{G}2\mathcal{W}$}
\newcommand{\RZSMark}{square*}
\newcommand{\RDMark}{star}
\newcommand{\GEEMark}{triangle}
\newcommand{\SMark}{o}
\newcommand{\DRZSMark}{diamond}
\newcommand{\DRDAMark}{diamond}

\title{Raptor Zonal Statistics: Fully Distributed Zonal Statistics of Big Raster + Vector Data [Pre-Print] }

\author{
 Samriddhi Singla   \\
 Computer Science and Engineering\\
    University of California, Riverside\\
  \texttt{ssing068@ucr.edu} \\
   \And
 Ahmed Eldawy \\
  Computer Science and Engineering\\
   University of California, Riverside\\
  \texttt{eldawy@ucr.edu} \\
  
 }

\begin{document}
\maketitle
\begin{abstract}
Recent advancements in remote sensing technology have resulted in petabytes of data in raster format. This data is often processed in combination with high resolution vector data that represents, for example, city boundaries. One of the common operations that combine big raster and vector data is the zonal statistics which computes some statistics for each polygon in the vector dataset.
This paper models the zonal statistics problem as a join problem and proposes a novel distributed system that can scale to petabytes of raster and vector data. The proposed method does not require any preprocessing or indexing which makes it perfect for ad-hoc queries that scientists usually want to run. We devise a theoretical cost model that proves the efficiency of our algorithm over the baseline method. Furthermore, we run an extensive experimental evaluation on large scale satellite data with up-to a trillion pixels, and big vector data with up-to hundreds of millions of edges, and we show that our method can perfectly scale to big data with up-to two orders of magnitude performance gain over Rasdaman and Google Earth Engine.
\end{abstract}
\section{Introduction}
\label{sec:introduction}

Advancements in remote sensing technology have led to a tremendous increase in the amount of remote sensing data. For example, NASA EOSDIS provides public access to more than 33~petabytes of Earth Observational data and is estimated to grow to more than 330~petabytes by 2025~\cite{EOSDIS}. European Space Agency (ESA) has collected over five petabytes of data within two years of the launch of the Sentinel-1A satellite and is expected to receive data continuously until 2030~\cite{ESA}. This big remote sensing data is available as raster data which is represented as multidimensional arrays. Many applications need to combine the raster data with vector data which is represented as points, lines, and polygons. This paper studies the {\em zonal statistics} problem which combines raster data, e.g., temperature, with vector data, e.g., city boundaries, to compute aggregate values for each polygon, e.g., average temperature in each city. This problem has several applications including the study by ecologists on the effect of vegetation and temperature on human settlement~\cite{JHB+07,JHS+11}, analyzing terabytes of socio-economic and environmental data~\cite{HMS17,HRM+15}, and studying of land use and land cover classification~\cite{saadat2011land}. It can also be used for areal interpolation~\cite{reibel2007areal} and to assess the risk of wildfires~\cite{thompson2016integrating}.

There exist many big spatial data systems which either process big {\em vector} data, e.g., SpatialHadoop~\cite{EM15}, GeoSpark~\cite{YSW15}, and Simba~\cite{XLY+16}, or big {\em raster} data, e.g., SciDB~\cite{SBZ+13}, RasDaMan~\cite{BDF+98}, and GeoTrellis~\cite{geotrellis-spark}. Unfortunately, these systems are not well-equipped to combine raster and vector data together and they all become very inefficient for the zonal statistics problem for big raster and vector data.

This paper models the zonal statistics problem as a {\em join problem} as it needs to find the pixels in the raster layer that overlap the polygons in the vector layer. We explain, with a support of a theoretical analysis, that existing approaches are analogous to two common join algorithms, namely, {\em index nested-loop join} and {\em hash join}. This analogy highlights the main limitations of these two algorithms as the data size increases. Therefore, we propose a novel distributed algorithm, termed {\em Raptor Zonal Statistics}, which resembles the {\em sort-merge join} for big raster and vector data. We show both analytically and experimentally that the proposed algorithm outperforms the baselines when the input data is very large.

Traditional methods to process the zonal statistics problem focused on either vectorizing the raster dataset~\cite{ZYG15} or rasterizing the vector data~\cite{HMS17}.
The first approach converts each pixel to a point and then runs a spatial join with polygons using a point-in-polygon predicate~\cite{ZYG15}. Finally, it groups the pixels by polygon ID and computes the desired aggregate function. This algorithm is analogous to an index-nested loop join algorithm which first builds an index for the vector dataset, i.e., polygons, and runs an index lookup for each pixel in the raster dataset. The drawback of this algorithm is the huge number of index lookups that makes the algorithm unpractical for big data.
The second approach rasterizes the vector data by converting each polygon to a raster (mask) layer with the same resolution of the input raster layer and then combines the two raster layers and compute the desired aggregate function~\cite{HMS17}. The one-to-one correspondence between the mask layer and raster layer makes this algorithm analogous to the hash join algorithm. This algorithm suffers from the huge size of the intermediate representation, i.e., hashtable, which increases with the input size.




This paper proposes a fully distributed algorithm, termed {\em Raptor Zonal Statistics} (RZS), which overcomes the limitations of the two baseline algorithms. As this paper shows, RZS is analogous to the {\em sort-merge join} algorithm and can provide highly efficient query processing with minimal overhead.
The proposed algorithm runs in three steps, namely,{\em intersection}, {\em selection} and {\em aggregation}. The intersection step partitions the vector data among machines using the R*-Grove algorithm~\cite{VE18} and computes an intermediate data structure, named {\em intersection file}, that contains a compact representation of the intersection of the vector and raster datasets. By only reading the metadata of the raster file, i.e., bounds and resolution, the intersection file is encoded in a way that allows the {\em selection} step to run in one scan over the raster data.
The second step, {\em Selection} has two phases, the first phase, {\em RaptorSplitGeneration} partitions both the intersection file and the raster data simultaneously using a new component called RaptorInputFormat. This component combines the record distribution of the vector data with the resolution of the raster data to produce fixed-size RaptorSplits that the machines can process in parallel while maintaining a balanced workload. The second phase of this step, {\em Raptor Data Processing} uses the intersection files to read pixel ranges from the raster files as defined by the {\em RaptorSplit}. The last step, {\em Aggregation} aggregates the pixel ranges read in the previous step to produce the required zonal statistics.
A previous work\footnote{Presented as a poster~\cite{Poster}} proposed a straight-forward parallelization of an efficient single-machine algorithm~\cite{eldawy2017large,SEA+19} which showed some promising results but was limited since it used the single-machine algorithm as a black-box. In this paper, we redesign the algorithm to make it fully distributed and we make a cost analysis to theoretically prove its efficiency over the baselines. Experiments on real datasets show that the proposed algorithm outperform all baselines for big data, including RasDaMan and Google Earth Engine, and show perfect scalability with large data and big clusters.

The rest of this paper is organized as follows.
Section~\ref{sec:related-work} covers the related work in literature.
Section~\ref{sec:background} provides a review of the concepts used in this paper.
Section~\ref{sec:Distributed Zonal Histogram} describes the proposed system, Raptor Zonal Statistics.
Section~\ref{sec:fullanalysis} provides a theoretical analysis and comparison of the proposed approach as well as the raster-based baseline.
Section~\ref{sec:experiments} provides an extensive experimental evaluation.
Finally, Section~\ref{sec:conclusion} concludes the paper. 

\section{Related Work}
\label{sec:related-work}

In this section we cover the relevant work in the literature. First, we give an overview of big spatial data systems and classify them according to whether they primarily target vector data, raster data, or both. After that, we cover the work that specifically targets the zonal statistics problem.

\subsection{Big Vector Data}
In this research direction, some research efforts aimed to provide big spatial data solutions for vector data types and operations. There are several systems in this category including SpatialHadoop~\cite{EM15}, Hadoop-GIS~\cite{AWV+13}, MD-HBase~\cite{NDA+13}, Esri on Hadoop~\cite{WPA+14}, GeoSpark~\cite{YSW15}, and Simba~\cite{XLY+16}, among others.
The work in this category covers
(1)~spatial indexes such as R-tree~\cite{EM15,YSW15,XLY+16}, Quad-tree~\cite{WPA+14,NDA+13}, and grid~\cite{EM15},
(2)~spatial operations such as range query~\cite{WPA+14,EM15,YSW15,XLY+16,NDA+13}, k nearest neighbor~\cite{EM15,YSW15,XLY+16,NDA+13}, spatial join~\cite{EM15,YSW15,XLY+16}, and computational geometry~\cite{ELM+13},
(3)~spatial data visualization including single-level and multilevel~\cite{EMJ16}, and
(4)~high-level programming languages such as Pigeon~\cite{EM14}.

Vector-based systems can support the zonal statistics problem by utilizing the index-nested loop join operation with the point-in-polygon predicate. Simply, it builds a spatial index for the vector layer to facilitate the point-in-polygon search. Then, it converts each pixel to a point and finds the containing polygon using the index. Shahed~\cite{EMA+15} further improves this query by building an aggregate Quad-tree index for the raster layer but it only supports rectangular regions while this paper considers complex polygons without the need to prebuild an index. When it comes to complicated polygons and high-resolution raster data, all these techniques become impractical.

\subsection{Big Raster Data}
Systems in this research direction focus on processing raster datasets which are represented as multidimensional arrays. Popular systems include SciDB~\cite{SBZ+13}, RasDaMan~\cite{BDF+98}, and GeoTrellis~\cite{geotrellis-spark}. The set of operations supported for raster datasets are completely different from those provided for vector datasets. They are usually categorized into four categories, namely, local, focal, zonal, and global operations~\cite{SC03}. Each operation operates on one or more multidimensional arrays and produce another multidimensional array possibly of a different size.

To support the zonal statistics operation for one polygon, a raster-based systems can apply the following algorithm which resembles a hash-join:
(1)~Rasterize the polygon by generating a raster layer of the same resolution of the input raster layer where a pixel has a value of one if it is inside the polygon. This layer serves as the hash table.
(2)~Apply a mask operation between the input raster and the rasterized layer which is a local operation.
(3)~Apply the desired statistics function on the resulting masked raster.
Another approach that may be used is to clip the raster layer using the polygon and aggregating the pixel values.
There are two drawbacks to these techniques. First, if the raster data has a very high resolution, the size of the rasterized polygon can be excessively large. Second, it has to process one polygon at a time and for vector datasets with thousands of polygons it will have to generate thousands of rasterized polygons, each with the resolution of the raster layer.

Wang {\em et al.}~\cite{wang2013parallel} proposed a parallel algorithm for rasterizing vector data which can be used to speed up the first step but it does not address the two limitations described above.

\subsection{Big Raster-Vector Combination}
Systems like PostGIS and QGIS~\cite{qgis} support vector and raster data on the user interface but they internally rely on two isolated libraries, one for each type of data. Therefore, they are still subject to the limitations described above.
One of the earliest works on combining raster and vector data is done in~\cite{peuquet1983hybrid}, which proposes a hybrid data structure to store both raster and vector data. It requires an offline preprocessing step that converts both datasets to an intermediate form before it performs any processing. The additional preprocessing step made it unattractive so our proposed approach is designed to work on the raster and vector data without any reformatting or preprocessing. Another work on querying raster and vector data~\cite{brisaboa} focuses on compact representation of raster in memory using a new tree-like data structure and performing range queries with vector data represented by an R-tree. On the other hand, the proposed approach does not require any index construction while achieving a better query performance.

\subsection{Zonal Statistics}
The {\em zonal statistics} problem is a basic problem that is used in several domains including ecology~\cite{JHB+07,JHS+11} and geography~\cite{HMS17,HRM+15}. However, there was only a little work in the query processing aspect of the problem. ArcGIS~\cite{arcgis} supports this query by first rasterizing the polygons dataset and then overlaying it with raster dataset which resembles a hash-join. Zhang {\em et al}~\cite{ZYG15,ZW14} solve the zonal statistics problem using an algorithm that resembles the indexed-nested loop join. It converts each pixel to a point and relies on GPUs to speed up the calculation. The drawback is that it has to load the entire raster dataset in GPU memory which is a very expensive operation and would be impractical for very large raster datasets. Zhao {\em et al}~\cite{zhao} aimed at increasing the performance of existing zonal statistics method using python in a shared memory multi-processor system. That method utilized threads by sending each thread a set of files of raster dataset and required rasterizing the polygon dataset as a separate step. 

Recent work in Terra Populus~\cite{HRM+15,HMS17} demonstrate the complexity of the problem on big raster and big vector datasets. The Scanline algorithm~\cite{eldawy2017large,SEA+19} was a first step in efficiently processing the zonal statistics problem by combining vector and raster data but it was limited to a single machine. In~\cite{Poster} (a poster), we tried a straight-forward parallelization of the Scanline approach that showed some performance improvement but it was limited as it used ScanLine as a black-box. In particular, it only parallelized the second phase that reads the raster data but still processes the vector data on a single machine which made it limited to small vector data.

In this paper, we propose a novel scalable sort merge join algorithm for processing the zonal statistics problem on big raster and vector data using MapReduce. It is different than the work described above in four ways.
\begin{enumerate*}
\item It leverages the MapReduce programming paradigm to scale out on multiple machines.
\item It is a fully distributed approach to the zonal statistics problem which allows it to scale to big raster and vector data.
\item It provides a novel mechanism for parallel task distribution that combines raster-plus-vector (Raptor) in one unit of work.
\item It can efficiently prune non-relevant parts in the raster layer to speed up the query processing.
\end{enumerate*}
\section{Review of Raster and Vector Data}
\label{sec:background}
This section provides a background on some relevant concepts from GIS and spatial databases, and the zonal statistics problem. Interested readers can refer to \cite{SC03} for more information. 

\subsection{Spatial Data Representation}
The two common representations of spatial data are {\em vector} and {\em raster} representations. The vector representation uses constructs like point, line, and polygon, and operations like intersect, union, and overlaps. The raster representation uses matrices as a common construct and the operations are all performed on these matrices. Each entry in the matrix is called a pixel. To map between pixels and geographic locations, two mappings are used, namely, {\em world-to-grid} (\wtg) and {\em grid-to-world} (\gtw). The {\wtg} mapping takes a coordinate in longitude and latitude and maps it to a position in the matrix and the {\gtw} mapping does the opposite. These mappings can also be used to map data between the vector and raster datasets. The algorithm in this paper relies mainly on these two mappings to map the computation between the two representations without having to convert one of them entirely to the other representation.

\subsection{Raster File Structure}
\label{sec:bg:raster}
A raster layer is modeled as very large dense matrix. One layer can typically contain trillions of entries. Most standard file structures, e.g., GeoTIFF and HDF5, partition this large matrix into smaller equi-sized {\em tiles}. Each tile is stored as one block and is typically small enough to load entirely in main memory. Tiles are identified by sequence numbers identifies, $t_{id}$, starting at zero. The file contains a lookup table that allows locating any tile efficiently. The data in all tiles can be stored in row-major or column-major order but they have to be the same. Finally, if a compression is applied, then the data in each tile is compressed separately to allow for decompression of a single tile. The proposed algorithm relies on this structure to achieve highly optimized solution while avoiding and prebuilt indexes.

\subsection{Zonal Statistics}
The input to the zonal statistics problem is a raster layer $r$, a vector layer $v$, and an accumulator $acc$. The vector layer $v$ consists of a set of polygons which are usually disjoint, e.g., city boundaries. The raster layer is a large two-dimensional matrix that contains remote sensing data, e.g., temperature values. The accumulator $acc$ is a user-provided function which takes pixel values, one at a time, and computes the statistics of interest. For example, one accumulator can compute the average value, while another one can compute the histogram over the spectrum of raster values. The output is a value for the accumulator for {\em each polygon} in the vector layer. For example, if $r$ represents the temperature in the world, $v$ represents the 50~US States, and $acc$ is an average accumulator, the output of this problem will be the average temperature for each state.
\subsection{Zonal Statistics on Raster DB}
\label{Sec:Arraydatabaseapproach}
The raster database approach focuses on processing raster data by representing them as multi-dimensional arrays and is used in~\cite{SBZ+13,BDF+98,geotrellis-spark,arcgis,HRM+15}. To solve the problem of zonal statistics, this approach follows a method called {\em clipping}. It runs in the following three steps.

\vspace{4pt} \noindent {\bf Step 1} rasterizes each polygon in the vector layer separately, by generating a mask layer of the same resolution of the input raster layer where a pixel has a value of one if it is inside the polygon and a zero if it is outside.

\vspace{4pt} \noindent {\bf Step 2} applies a masking operation between the input raster and the rasterized mask which is a local operation.

\vspace{4pt} \noindent {\bf Step 3} computes the desired aggregate function by aggregating all the values in the masked layer.

\subsection{Zonal Statistics on Vector DB}
\label{sec:Spatialdatabaseapproach}
The vector database approach focuses on processing vector data by representing each feature in terms of a list of coordinates associated with a set of attributes. A feature can be a point, a line or a polygon. This approach is used in \cite{EM15,YSW15,XLY+16}. To solve the problem of zonal statistics, this approach follows a method called {\em point-in-polygon}. It runs in the following three steps.

\vspace{4pt} \noindent {\bf Step 1} builds a spatial index for the vector layer to facilitate the point-in-polygon query. That is, given a query point, find the containing polygon.

\vspace{4pt} \noindent {\bf Step 2} converts each pixel to a point and runs an index lookup to find the containing polygon, if any.

\vspace{4pt} \noindent {\bf Step 3} groups the points by their containing polygon ID and runs the desired aggregate function on each group.

\subsection{Single-machine ScanLine Method}
\label{sec:scanline}
The scan-line method~\cite{eldawy2017large} is the state-of-the-art algorithm for computing zonal statistics on a single machine. It runs in the following three steps.

\vspace{4pt} \noindent {\bf Step 1} calculates the Minimum Bounding Rectangle (MBR) of the input polygon(s) and maps its two corners from world to grid coordinates using the {\wtg} mapping. These two corners help in identifying the lower and upper rows in the grid coordinates that define the range of scan lines to process.

\vspace{4pt} \noindent {\bf Step 2} computes the intersections of each of the scan lines with the polygon boundaries. It converts each scan line from grid coordinates to world coordinates using {\gtw} mapping, and stores their $y$-coordinates in a sorted list. Each polygon is scanned for its corresponding range of scan lines, which are then used to compute intersections with the polygon. These intersections are then sorted by their $x$-coordinates for each scan line.

\vspace{4pt} \noindent {\bf Step 3} finds the pixels that lie inside the polygons and process them. It maps the $x$-coordinates of the intersections from world to grid coordinates and accumulates the corresponding pixel values. For multiple polygons, all intersections in one row are processed before moving to the next row.  

This approach requires a minimal amount of intermediate storage for the intersection points. It also minimizes disk IO by scanning the raster data exactly once and by reading only the pixels that overlap the polygons.

\section{Raptor Zonal Statistics (RZS)}
\label{sec:Distributed Zonal Histogram}

\begin{figure*}[t]
\centering
\includegraphics[width=\textwidth]{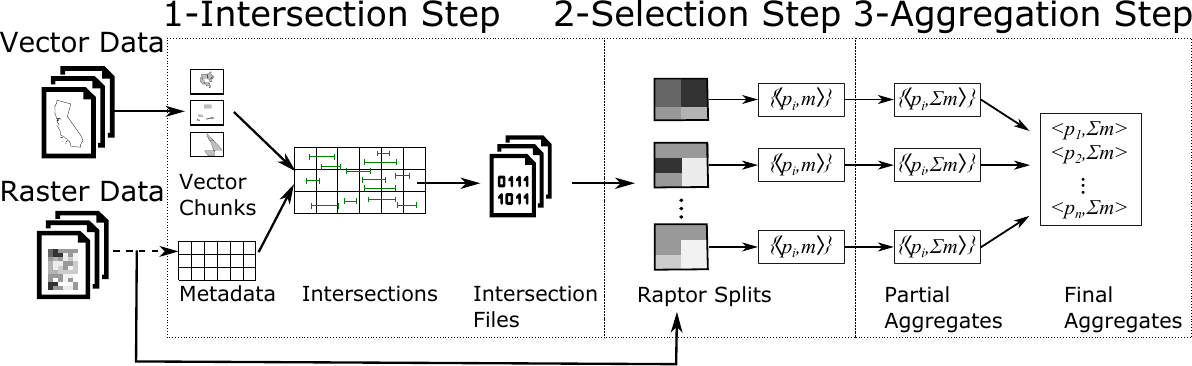}
\caption{Overview of the Raptor join algorithm}
\label{figure:systemoverview}
\end{figure*}

In this section, we describe the proposed {\em Raptor Zonal Statistics} (RZS) algorithm that follows a {\em sort-merge join} approach to join raster and vector datasets and answer the zonal statistics query. By observing the analogy between the two baseline approaches and the two join algorithms, hash join and index nested loop, the reader can see why neither of them scales to big raster and vector data. The hash join does not scale due to the huge size of the hashtable since there must be a unique key for each pixel. The indexed nested loop join does not scale either due to the overwhelming cost of index lookups or the large size of the index. This leaves us with the sort-merge join approach which is usually more efficient than both as it requires one scan over the data. However, it requires an expensive sorting phase which overweighs the saving in the join phase. Therefore, the sort-merge join algorithm is typically used only when the input datasets are already sorted.

The key idea of the proposed algorithm is that we exploit the internal structure of the raster data and produce an intermediate compact representation of the vector data, called {\em intersection file}, which perfectly matches the order of the raster data. Furthermore, to produce the intersection file, we only need to process the vector layer and the {\em metadata} of the raster layer which means that the raster dataset needs to be scanned once. Finally, the intersection file is generated and stored in a distributed fashion which allows the proposed algorithm to be parallelized over a cluster of machines.

This algorithm runs in three steps, namely, {\em intersection}, {\em selection}, and {\em aggregation}. The {\em intersection} step computes the intersection file which captures the intersections between the vector and raster data and sorts these intersections to match the raster data. The {\em selection} step uses the intersection file to read the pixels in the raster layer that intersect each polygon in the vector layer. Finally, the {\em aggregation} step groups the pixel values by polygon ID and computes the desired aggregate function, e.g., average. The details of the three steps are given below.

\subsection{Step 1: Intersection}

This step runs as a map-only job and is responsible for {\em intersections file generation}. It takes as input the vector layer and the {\em metadata} of the raster layer and computes a common structure, called {\em intersection file}, which is stored in the distributed file system to be used in the selection step. The vector layer consists of a set of polygons each represented as a list of straight line segments. The metadata of the raster layer consists of the dimensions (number of rows and columns), two affine matrix transformations {\gtw} and {\wtg} as described in Section~\ref{sec:background}, and the size of each tile in the raster layer, i.e., number of rows and column, and is only a few kilobytes.

The input vector layer is partitioned into fixed-size chunks, 128~mb by default, and each chunk is assigned to one task. While any partitioning technique works fine, we employ the R*-Grove partitioning technique~\cite{VE18} which maximizes the locality of partitions while ensuring load balance. In Spark, each chunk can be processed using the {\tt mapPartitions} transformation while in Hadoop it can be processed as a map task using the {\tt run} method.

For each chunk, this step computes all the intersections between each row in the raster layer and each segment of each polygon in the vector layer. To compute these intersections, each line segment is mapped to the raster layer using the {\wtg} transformation to find the range of rows that intersect. Then, it is a simple constant-time computation to find the $x$-coordinate of the intersection. Since we only need to know the intersection at the pixel level, the intersection is mapped to the raster space and the integer coordinate of the pixel is computed. We record the intersection as the triple $\langle p_{id},x,y\rangle$ where $p_{id}$ is the polygon ID to which the segment belongs and $(x,y)$ is the coordinate of the intersection in the raster layer. All these triplets are kept in memory and can be spilled to disk if needed.

After all the intersections are computed, we run a sorting phase which sorts the intersections lexicographically by $(t_{id},y,p_{id},x)$\footnote{the sorting order becomes $(t_{id},x,p_{id},y)$ if the raster file is stored in column-major order}, where $t_{id}$ is the grid tile ID that contains the intersection. Notice that the grid tile ID does not have to be explicitly stored since it can be computed in a constant time for each intersection using the metadata of the raster layer.

\begin{figure}[t]
\centering
\includegraphics[width=0.5\columnwidth, scale = 0.7]{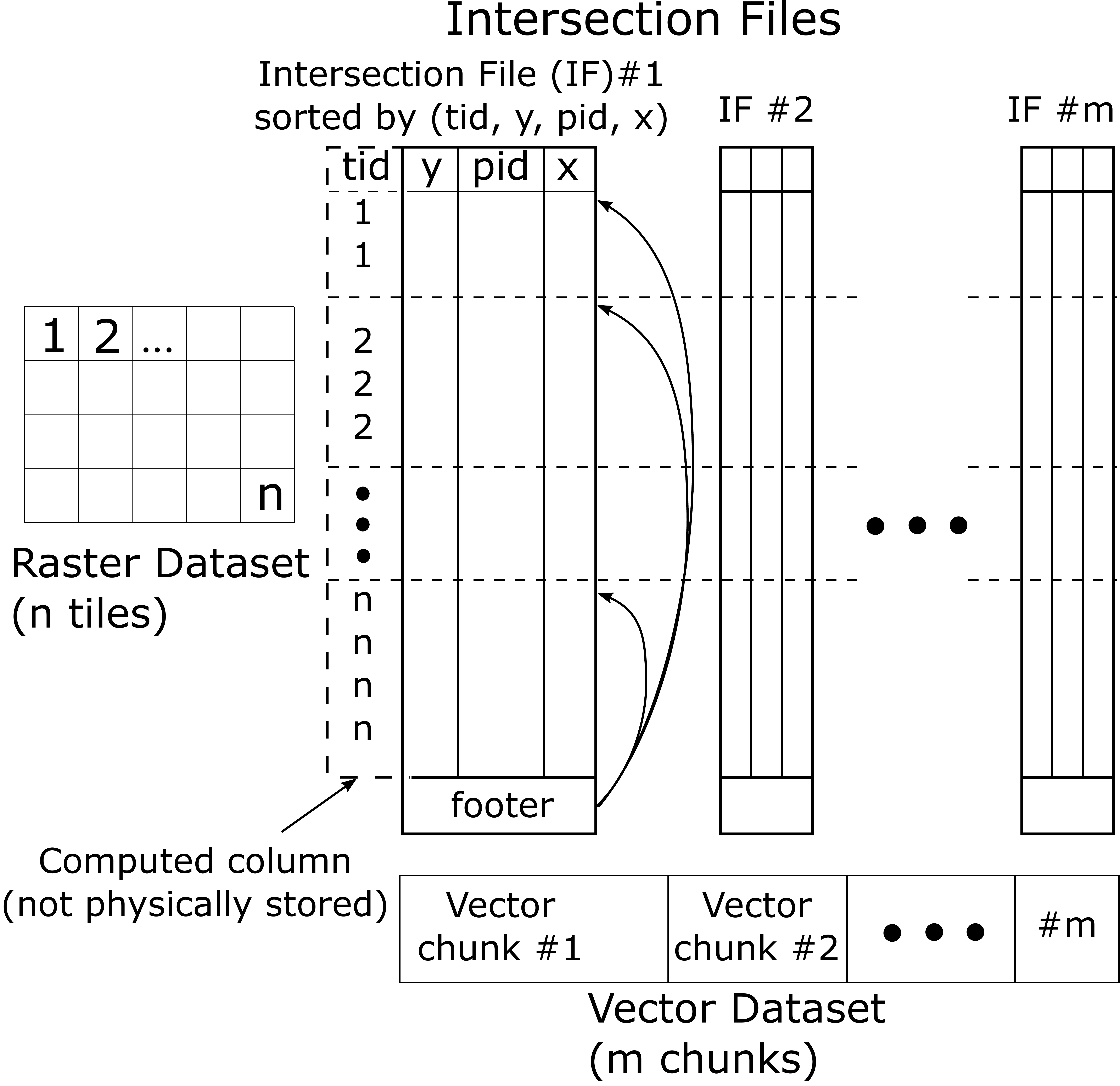}
\caption{Intersection file structure}
\label{figure:intersectionfile}
\end{figure}

Finally, the sorted intersections are stored in the intersection file as illustrated in Figure~\ref{figure:intersectionfile}. The figure shows multiple intersection files, one for each chunk in the vector data. Since each chunk is processed on a separate machine, the intersection files are computed and written in a fully distributed manner. In each intersection file, the intersections are stored in the sorted order $(t_{id},y,p_{id},x)$. As mentioned earlier, $t_{id}$ is not physically stored to save space and is computed as needed. In addition, we append a footer to each intersection file which stores the list of polygon IDs that appear in this file and a pointer to the first intersection in each grid tile ID $t_{id}$. This imposes a logical partitioning of the intersection file by $t_{id}$ as illustrated by the dotted lines in the figure. This footer is only a few kilobytes for a hundred megabyte file and does not impose any significant overhead.

\subsection{Step 2 : Selection}
\label{sec:rj:selection}

This step uses the intersection files produced in the first step to select all the pixels that are contained within each polygon. The input to this step is the intersection files and the raster layer while the output is a set of pairs $\langle p_{id}, m\rangle$ where $p_{id}$ is the ID of the polygon and $m$ is the measurement in the pixel. Notice that these values are pipelined to the next step and are not physically stored as there is typically hundreds of billions of these pairs.

This step runs in two phases, namely \emph{Raptor Split Generation} and \emph{Raptor Split Processing}. The first phase creates a list of tasks that is distributed among machines to perform the parallel computation. The second phase processes the raster files and intersection files as defined in by the Raptor splits.

\vspace{4pt} \noindent {\bf Raptor Split Generation}\\
The \emph{Raptor Split Generation} phase generates \emph{Raptor Splits} using the \emph{RaptorInputFormat}. In both Spark and Hadoop, the InputFormat is the component that splits the input file(s) into roughly equi-sized splits to be distributed on the worker nodes. These splits are mapped one-to-one to mappers. Therefore, each split defines a unit of work called map task. A corresponding {\em record reader} uses the split to extract key-value pairs that are sent to the map function for processing. Since our unit of work is a combination of raster plus vector data, we define the new {\em RaptorInputFormat}, {\em RaptorSplit}, {\em RaptorRecordReader}, and {\em RaptorObject}, described as follows. Starting with the smallest one, the RaptorObject contains vector chunk ID $[1,m]$, a raster tile ID $[1,n]$. The RaptorObject defines the smallest unit of work done by the map function.
The RaptorSplit stores a vector chunk ID and a {\em range} of tile IDs. The RaptorSplit defines the task given to each mapper. We can control the amount of work given to each machine by adjusting the number of tiles in the RaptorSplit. The RaptorRecordReader takes one RaptorSplit and generates all the RaptorObjects that it represents. Finally, the RaptorInputFormat takes all the input to the problem, i.e., the raster dataset and all intersection files, and produces a list of RaptorSplits that define the map tasks given to worker nodes. The information needed to create the RaptorSplits is all contained in the {\em footers} of the intersection files. Those footers contain the list of tile IDs that intersect each vector chunk. We use this information to create RaptorSplits that contain those tiles that overlap each vector chunk. Effectively, this step prunes all the tiles that do not contribute to the answer.

As an additional optimization, if the raster dataset is stored in multiple files, this phase will limit the range of tile IDs in each RaptorSplit to only one file. This ensures that each machine in this step will need to process only one file. On the other hand, if the raster layer is stored in one big file, we can still split that file among multiple machines for efficient processing.

\vspace{4pt} \noindent {\bf Raptor Data Processing}\\
This phase takes a RaptorObject, which contains a vector chunk ID and a tile ID, and generates a set of pairs $\langle p_{id},m\rangle$ where $p_{id}$ is a polygon ID and $m$ is a measurement in a pixel contained in that polygon. In Spark, this phase can be implemented as a {\tt flatMap} transformation and in Hadoop it can be implemented as part of the {\tt map} function. This phase starts by loading the raster tile indicated by the tile ID $t_{id}$ in the input RaptorObject. One tile is typically small enough to fit in main memory so the entire tile is loaded as a two-dimensional array. After that, it opens the intersection file indicated in the RaptorObject. The footer is loaded to locate the first intersection in the tile $t_{id}$. Then it scans the intersections in the intersection file. Each pair of intersections for one polygon indicate a range of pixels contained inside that polygon. The corresponding range of pixels is scanned from the raster file and a set of pairs $\langle p_{id},m \rangle$ is produced where $p_{id}$ is defined in the intersection and the $m$ values are defined in the loaded grid tile.

\subsection{Step 3: Aggregation}
In this last step, the set of pairs $\langle p_{id}, m\rangle$ are aggregated to produce final aggregate values $\langle p_{id}, \sum m \rangle$, where $\sum$ is any associative and commutative aggregate function. As in most distributed systems, the aggregate function is computed in two phases, {\em partial} and {\em final} aggregates. The partial/final computation supports any function that is both associative and commutative which covers a wide range of aggregate functions. The {\em partial} aggregates are computed locally in each machine while the {\em final} aggregates are computed by combining all the partial aggregates for the same group in one machine. In Spark, this can be implemented as a {\tt aggregateByKey} transformation. In Hadoop, this can be implemented as a pair of {\tt combine} and {\tt reduce} functions. 

\section{Full Analysis}
\label{sec:fullanalysis}
This section shows the full analysis including computation and disk cost for the three algorithms, RDA, VDA, and RZS. This analysis only considers steps~1\&2 of the three algorithms since step~3 is identical for all of them. This part is done on the basis of the size of input raster and vector datasets. Table~\ref{tab:fullsymbols} summarizes the parameters used throughout the analysis.

\begin{table}[h]
\begin{center}
\caption{Parameters for Cost Estimation}
\flushleft
\begin{center}
\normalsize
\begin{tabular}{|c|l|}
\hline
Symbol & Meaning\\
\hline
$r$ & Number of rows in raster \\
$c$ & Number of columns in raster\\
$w_t$ & Tile width in pixels \\
$h_t$ & Tile height in pixels \\
$p$ & Pixel size in degrees\\
$n_p$ & Number of polygons\\
$n_s$ & Number of line segments in all polygons\\
$\overline{n_s}=\frac{n_s}{n_p}$ & Average number of line segments per polygon\\
$w_p$ & Average polygon width in degrees\\
$h_p$ & Average polygon height in degrees\\
$h_s$ & Average line segment height in degrees\\
$I$ & Input size in bytes\\
$B$ & HDFS block size in bytes\\
$C$ & Chunk size. Number of polygons in intersection file\\
\hline
\end{tabular}
\end{center}
\label{tab:fullsymbols}
\end{center}
\end{table}

\subsection{Raster Database Approach (RDA)}
\label{sec:Analysis:ADA}
The RDA algorithm scans the polygons and for each polygon it clips the overlapping portion of the raster layer. Finally, it scans all the clipped pixels and aggregates them. The {\em clipping} operation is implemented efficiently by sorting all line segments in the polygon. For $n_p$ polygon with an average of $\overline{n_s}$ segments per polygon, the cost of the sorting operation is:
\begin{equation}
    T_s=n_p\cdot \overline{n_s} \log{\overline{n_s}}
\end{equation}

After that, the algorithm needs to locate the clipped pixels in the raster file. Since the input raster file is stored as tiles of size $w_t\times h_t$ pixels, it needs to read the entire tiles that overlap the mask layer. On average, each polygon will overlap with $n_t$ tiles as given below.
\begin{equation}
    n_t=\left\lceil \frac{w_p}{w_t\cdot p}\right\rceil\times 
        \left\lceil \frac{h_p}{h_t\cdot p}\right\rceil
\end{equation}

Finally, the total time for the {\em clipping} step, $T_c$ is computed as follow.
\begin{equation}
    T_c = n_p\cdot n_t \cdot w_t \cdot h_t
\end{equation}

Hence, the total computation time for RDA,
\begin{equation}
    T_{RDA}=T_s+T_c
    \label{eqn:t_rda}
\end{equation}

\subsection{Vector Database Approach (VDA)}
This approach indexes the vector dataset and, for each pixel in the raster dataset, it checks if that pixel is in the polygons or not. This requires a scan through the whole raster dataset, while searching which polygon contains that pixel. If the pixel lies in any polygon, it simply adds it to the set of values and aggregates them at the end. It takes at most $T_{VDA}$ time computed as
\begin{equation}
    T_{VDA} = n_s \log n_s + c\cdot r\cdot \log n_s 
    \label{eqn:t_vda}
\end{equation}

$n_s\log n_s$ represents the time needed to build an efficient index, e.g., k-d tree, to index the polygons. The term $c\cdot r\cdot \log n_s$ represents the lookup phase which performs an index lookup, with cost $\log n_s$ for each pixel in the raster layer, with dimensions $c\times r$.

The disk I/O, $D_{VDA}$, for this approach requires to read the vector dataset to index it and then read and write the index to and from the disk. While the index can be kept in memory for small vector data, this paper focuses more on the case of large data, thus, we assume that the vector data cannot fit in main memory. Apart from this, the approach requires to read the entire raster layer, hence
\begin{equation}
    D_{VDA} =  c \cdot r + I
    \label{eqn:d_vda}
\end{equation}
This approach is analogous to {\em index-nested loop join}, where the vector data is indexed and then then the point-in-polygon query is run as in a nested loop.

\subsection{Raptor Zonal Statistics (RZS)}
\label{sec:analysis:rzs}

Below, we estimate the computation time per chunk in RZS which is split into an intersection and selection steps. 

{\bf Intersection step:}
For each vector chunk, this step is responsible for computing the intersections of the polygons in that chunk with the raster layer. It takes as input the vector polygons and the raster metadata. This step simply scans all line segments and computes the intersections with all scan lines. Let $\overline{n_i}$ be the average number of intersections per intersection file. Then,
\begin{equation}
    \overline{n_i} = n_s \frac{h_s}{p}
    \label{eqn:k}
\end{equation}
where $h_s$ is the average height of a line segment in {\em degrees}. This makes $\frac{h_s}{p}$ equal to the average height of a line segment in {\em pixels}. If $h_s$ is not know exactly, we can approximate it based on a polygon height and the average number of segments per polygon. On an average polygon, the segments would travel the height of the polygon twice to make a closed polygon. Thus,
\begin{equation}
    h_s=\frac{h_p}{2\cdot \overline{n_s}}
\end{equation}

After the intersections are computed, we need to sort them in the intersection file. The average time per chunk for the intersection step $\overline{T_I}$, including sorting, is given as below:
\begin{equation}
    \overline{T_I} = \overline{n_i} + \overline{n_i} \log{\overline{n_i}}
\end{equation}
where $\overline{n_i} \log{\overline{n_i}}$ is the sorting time.

{\bf Selection step:}
This step reads the intersections from the {\em intersection files} and then reads the raster values between each pair of intersections to compute the required {\em zonal statistics}. Similar to RDA, we need to account for the raster file structure while the smallest access unit is a tile. The average number of tiles that overlap one chunk $\overline{n_t}$ is calculated as follow:
\begin{equation}
 \overline{n_t} = \lceil \frac{w_c}{w_t} \rceil\times \lceil \frac{h_c}{h_t} \rceil
\end{equation}

Finally, the average cost of the selection step per chunk $\overline{T_S}$ is calculated as the total number of pixels read from the raster file.
\begin{equation}
    \overline{T_S}=\overline{n_t}\times w_t \times h_t
\end{equation}
And the average cost to process one chunk in the RZS algorithm is $\overline{T_{RZS}}=\overline{T_I}+\overline{T_S}$. Given $n_c$ chunks, the expected processing cost of the RZS algorithm is:
\begin{equation}
    T_{RZS}=n_c\cdot \overline{T_{RZS}}
    \label{eqn:t_rzs}
\end{equation}

\subsection{Discussion}
Comparing Equations~\ref{eqn:t_rda},~\ref{eqn:t_vda}, and~\ref{eqn:t_rzs}, we observe that the total computation time for RZS is the least, since it is limited by the time taken to read the raster. It only has to pay the overhead of creating and sorting the intersection files, with total size $K$, which is proportional to the raster layer resolution $p$ (Equation~\ref{eqn:k}).
On the other hand, RDA suffers from the huge cost of creating the mask layers (hashtables) which have a total size that is proportional to the squared resolution $p^2$ in Equation~\ref{eqn:t_rda}. It also requires reading portions of the raster data multiple times if they overlap multiple polygons. Hence, the total cost of the RDA approach is several times the size of the raster layer. Finally, the computation cost of VDA is prohibitive as it requires running a large number of index lookups which is equivalent to processing the raster layer multiple times.

\section{Experiments}
\label{sec:experiments}

This section provides an experimental evaluation of the proposed algorithm, Raptor Zonal Statistics (RZS). We compare the distributed Raptor Zonal Statistics algorithm to the single-machine Scanline Method, Rasdaman (a RasterDB approach) and Google Earth Engine (GEE). We show that the proposed RZS is much faster than Rasdaman and has performance comparable with Google Earth Engine. We evaluate them on real data and also show the effect of various design decisions on the Raptor Zonal Statistics. 
\begin{table}
\caption{Vector and Raster Datasets}
\centering
\normalsize
Vector datasets\vspace{4pt}\\
\footnotesize
\begin{tabular}{|l|r|r|r|r|r|r|}
\hline
Dataset    & $n_p$ & $n_s$ & $\overline{w_p}$ & $\overline{h_p}$ & File Size $I$\\
\hline
Counties   & 3k    & 52k   & 0.82             & 0.51             & 978~KB   \\
States     & 49    & 165k  & 12.18            & 4.28             & 2.6~MB   \\
Boundaries & 284   & 3.8m  & 18.61            & 8.18             &  60~MB   \\
TRACT      & 74k   & 38m   & 0.096            & 0.068            & 632~MB   \\
ZCTA5      & 33k   & 53m   & 0.19             & 0.15             & 851~MB   \\
Parks      & 10m   & 336m  & 0.0067           & 0.0043           & 8.5~GB   \\
\hline
\end{tabular}

\vspace{10pt}

\normalsize
Raster datasets\vspace{4pt}\\
\footnotesize

\begin{tabular}{|l|c|c|r|}
\hline
Dataset    & image size $c\times r$     & tile size $w_t\times h_t$ & resolution $p$ \\
\hline 
glc2000    &    $40,320 \times 16,353$  & $128    \times 128$       & 0.0089  \\
MERIS      &   $129,600 \times 64,800$  & $256    \times 256$       & 0.0027  \\
US-Aster   &   $208,136 \times 89,662$  & $208136 \times 1  $       & 0.00028 \\
Tree cover & $1,296,036 \times 648,018$ & $36001  \times 1  $       & 0.00028 \\
\hline
\end{tabular}

\label{tab:datasets}
\end{table}

Section~\ref{sec:experiments:setup} describes the setup of the experiments, the system setup and the datasets used. 
Section~\ref{sec:experiments:comparison} provides a comparison of the proposed RZS, Scanline method, Rasdaman and Google Earth Engine based on the total running time. It also discusses the vector and raster dataset ingestion time for each of these methods. 
Section~\ref{sec:experiments:verification} gives a verification of the proposed cost model for RZS and RDA methods.
Section~\ref{sec:experiments:applications} discusses two applications where the proposed RZS algorithm has been used.
Section~\ref{sec:experiments:tuning} shows the effect of various parameters on the total running time of RZS. The parameters include the size of vector chunks, compression of intersection files and effect of partitioning the vector data.

\subsection{Setup}
\label{sec:experiments:setup}
We run all the experiments on a Amazon AWS EMR cluster with one head node and $19$ worker nodes of type m4.2xlarge with $2.4$ GHz Intel Xeon $E5-2676$ v3 processor, $32$ GB of RAM, up to $100$ GB of SSD, and 2$\times$8-core processors. The methods are implemented using the open source GeoTools library 17.0.

In all the techniques, we compute the four aggregate values, minimum, maximum, sum, and count. We measure the end-to-end running time as well as the performance metrics which include reading both datasets from disk and producing the final answer.
Table~\ref{tab:datasets} lists the datasets that are used in the experiments along with their attributes using the terminology in Table~\ref{tab:fullsymbols}. All these datasets are publicly available as detailed in~\cite{Poster}. The {\em boundaries} and {\em parks} datasets cover the entire world while the other four datasets cover only the continental US. All raster datasets cover the entire world except {\em US-aster} which only covers the continental US.

RZS is implemented in Hadoop 2.9. We chose to implement it in Hadoop rather than Spark for two reasons. First, it was simpler because Hadoop can easily support custom input format, i.e., the RaptorInputFormat while Spark does not have a specialized method to define a new input format; it just reuses Hadoop's InputFormat architecture. Second, Spark is optimized for in-memory processing while RZS is a disk-intensive query that does not have a huge memory footprint.

For RDA, we used Rasdaman 10.0 running on a single machine since the community version does not support distributed processing. We also used Google Earth Engine (GEE) which runs on Google cloud engine. GEE is still experimental and is currently free to use. The caveat is that it is completely opaque and we do not know which algorithms or how much compute resources are used to run queries. Therefore, we run each operation on GEE 3-5 times at different times and report the average to account for any variability in the load. For large vector data, we hit the limit of GEE of 2GB vector file. To work around it, we split the file into 2GB smaller files, run on each file separately, and add up the results. All the running times are collected as reported by GEE in the dashboard.

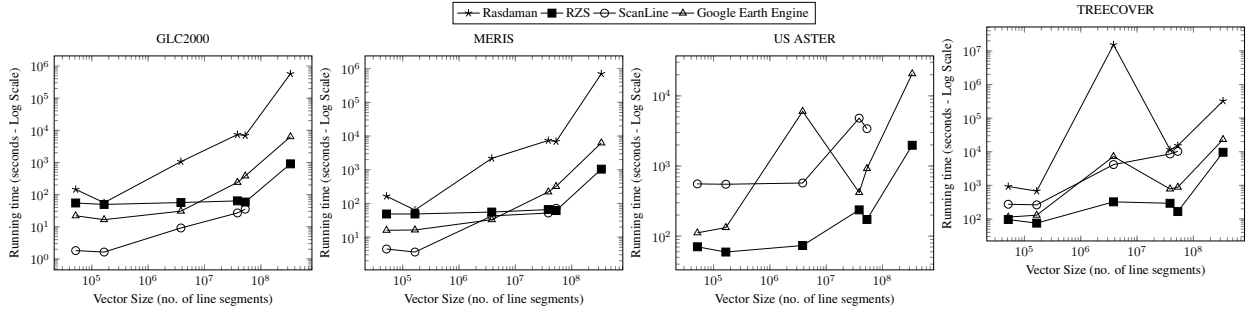
\begin{figure*}[ht]
\begin{minipage}[b]{0.245\linewidth}
\centering
\pgfplotstableread{ 
VectorSize  Rasdaman    RZS ScanLine    GoogleEarthEngine
51638       145.4109702	54.92	1.82	22
165186      56.08144212	49.55	1.64	16.66666667
3817412     1060.837246	56.71	9.18	30.66666667
38467094    7360.893012	63.92	26.92	240
52894188    6869.52474	58.84   35.36   380
336145640   5.73E+05    905.35  nan     6320
}\testdata

\begin{tikzpicture}[scale=0.5]
    \begin{loglogaxis}
    [legend style={at={(2.25,1.2)}, anchor=center,legend columns=4},
    title={GLC2000},
        xlabel={Vector Size (no. of line segments)},
        ylabel={Running time (seconds - Log Scale)}]
    \addplot[mark=\RDMark, mark size=3]
    table[x index=0,y index=1]{\testdata};
       \addlegendentry{Rasdaman}
    \addplot[mark=\RZSMark, mark size=3]
    table[x index=0,y index=2]{\testdata};
       \addlegendentry{RZS}
    \addplot[mark=\SMark, mark size=3]
    table[x index=0,y index=3]{\testdata};
       \addlegendentry{ScanLine}
    \addplot[mark=\GEEMark, mark size=3]
    table[x index=0,y index=4]{\testdata};
    \addlegendentry{Google Earth Engine}
    \end{loglogaxis}
    \end{tikzpicture}   
\end{minipage}
\begin{minipage}[b]{0.245\linewidth}
\centering
\pgfplotstableread{ 
VectorSize  Rasdaman	RZS	    ScanLine GoogleEarthEngine
51638	    164.4613039	48.68	4.45	 16
165186	    63.66582084	48.73	3.62	 16.33333333
3817412	    2179.372624	55.60	42.73	 33
38467094	7397.219183	65.78	52.08	 220
52894188	6888.428213	61.63	71.53    320
336145640	7.06E+05    1043.68 nan      6260 
}\testdata

\begin{tikzpicture}[scale=0.5]
    \begin{loglogaxis}
    [legend style={at={(1,0)}, anchor=center,legend columns=3},
    title={MERIS},
        xlabel={Vector Size (no. of line segments)},
        ylabel={Running time (seconds - Log Scale)}]
    \addplot[mark=\RDMark, mark size=3]
    table[x index=0,y index=1]{\testdata};
    \addplot[mark=\RZSMark, mark size=3]
    table[x index=0,y index=2]{\testdata};
     \addplot[mark=\SMark, mark size=3]
    table[x index=0,y index=3]{\testdata};
    \addplot[mark=\GEEMark, mark size=3]
    table[x index=0,y index=4]{\testdata};
    \end{loglogaxis}
    
\end{tikzpicture}   
\end{minipage}
\begin{minipage}[b]{0.245\linewidth}
\centering
\pgfplotstableread{ 
VectorSize  Rasdaman    RZS	    ScanLine  GoogleEarthEngine
51638	    nan	        70.62	555.99	  111.6666667
165186	    nan	        59.46	549.98	  132.6666667
3817412	    nan	        73.67	572.81	  6013.666667
38467094	nan	        237.29	4808.63	  420
52894188	nan	        173.14	3407.19	  920
336145640	nan	        1977.13	nan	      20660
}\testdata
\begin{tikzpicture}[scale=0.5]
    \begin{loglogaxis}
    [    legend style={at={(1,0)}, anchor=center,legend columns=3},
    title={US ASTER},
        xlabel={Vector Size (no. of line segments)},
        ylabel={Running time (seconds - Log Scale)}]
    \addplot[mark=\RDMark, mark size=3]
    table[x index=0,y index=1]{\testdata};
    \addplot[mark=\RZSMark, mark size=3]
    table[x index=0,y index=2]{\testdata};
     \addplot[mark=\SMark, mark size=3]
    table[x index=0,y index=3]{\testdata};
    \addplot[mark=\GEEMark, mark size=3]
    table[x index=0,y index=4]{\testdata};
    \end{loglogaxis}
\end{tikzpicture}   
\end{minipage}
\begin{minipage}[b]{0.245\linewidth}
\centering
\pgfplotstableread{ 
VectorSize  Rasdaman	RZS	            ScanLine	GoogleEarthEngine
51638	    925.6752756	96.02487806	    274.21	    113.6666667
165186      675.7143731 75.26464908	    262.99	    129.6666667
3817412     152E+05   322.898113727   4153.31     7233
38467094    11508.87379 293.143977347   8599.146407 780
52894188	15261.39102 166.318463914   10270.86765 880
336145640	3.25E+05    9660.287384438	nan	        23160         
}\testdata
\begin{tikzpicture}[scale=0.5]
    \begin{loglogaxis}
    [legend style={at={(1,0)}, anchor=center,legend columns=3},
    title={TREECOVER},
        xlabel={Vector Size (no. of line segments)},
        ylabel={Running time (seconds - Log Scale)}]
    \addplot[mark=\RDMark, mark size=3]
    table[x index=0,y index=1]{\testdata};
    \addplot[mark=\RZSMark, mark size=3]
    table[x index=0,y index=2]{\testdata};
     \addplot[mark=\SMark, mark size=3]
    table[x index=0,y index=3]{\testdata};
    \addplot[mark=\GEEMark, mark size=3]
    table[x index=0,y index=4]{\testdata};
     \end{loglogaxis}
\end{tikzpicture}   
\end{minipage}

\caption{Comparison of total running time of RZS, Scanline, Google Earth Engine and Rasdaman}
\label{figure:comparison}
\end{figure*}

\subsection{Overall Execution Time}
\label{sec:experiments:comparison}

This parts compares RZS, Scanline, Rasdaman, and GEE based on the end-to-end execution time. This experiment is run for all the combinations of vector and raster datasets shown in Table~\ref{tab:datasets}, and its results can be seen in Figure~\ref{figure:comparison}. We omit results for Rasdaman for parks dataset and the combination of boundaries and treecover datasets as it takes more than forty-eight hours to compute them. All experiments on RZS run on a cluster of 20 machines except the TreeCover dataset which we run on 100 nodes due to its huge size. 

As can be observed from Figure~\ref{figure:comparison}, the proposed distributed RZS algorithm is orders of magnitude faster than Rasdaman for all combinations of raster and vector datasets. RZS is up-to two orders of magnitude faster than GEE and twice as fast for the largest input (Parks$\bowtie$TreeCover).

{\bf Rasdaman} failed to ingest the US-Aster file due to its huge size (48GB as a single BigGeoTIFF file). In addition to being a single machine, Rasdaman does not scale due to using the RDA method which scans the polygons, clips the raster layer for each polygon, and aggregates the clipped values. Based on the overlap of polygons and raster tiles, the same tile could be read tens of times based on overlapping polygons. RZS overcomes this problem by generating intersection files that are ordered based on the raster file structure to ensure a single scan of the raster file.


When compared to {\bf GEE} for large datasets, RZS is at least 2x faster and up-to two orders of magnitude faster. In particular, the speedup of RZS is much higher for large vector datasets which indicates that GEE uses raster-based methods which focus on large raster datasets but ignore the size of the vector data. We further confirm this finding using our cost model in the next part. While GEE is faster for small vector datasets, this is due to the overhead of using Hadoop for small inputs. Indeed, the single machine Scanline~\cite{eldawy2017large} algorithm is an order of magnitude faster than both for these small datasets, e.g., GLC2000 and MERIS for small vector data. Therefore, if we want to be always faster than GEE, we just need to switch to Scanline for small datasets but we leave this for a future work.
GEE is a free tool (for now) but the knowledge of how it implements the zonal statistics operation\footnote{reduceRegions function in GEE} is not public. Also, the amount of resources available to users at any time can vary and this is why we report the average of three runs.

When compared to the single-machine {\bf Scanline} we observe two orders of magnitude speedup of RZS due to the parallel implementation. We also observe that Scanline cannot scale to large vector data due to the limitations of the intersection step which runs on a single machine and that is why the straight-forward parallelization of Scanline on Hadoop~\cite{Poster} did not scale either.

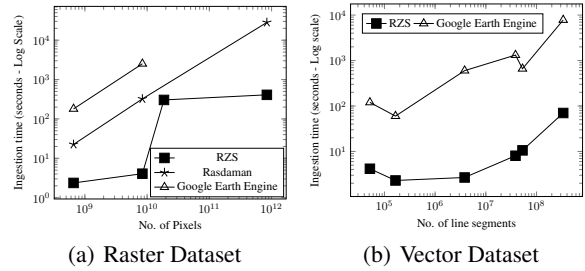
\begin{figure}
    \centering
    \subfigure[Raster Dataset]{\pgfplotstableread{ 
Raster     RZS       Rasdaman    GEE
657881190    2.372     22.54       180
8398080000      4.068     323.2       2520
18661890032  304.059   nan         nan
839854656648  409.672   28224       nan
}\testdata

\begin{tikzpicture}[scale=0.45]
    \begin{loglogaxis}
    [legend style={at={(0.99,0.01)}, anchor=south east,legend columns=1},
        xlabel={No. of Pixels},
        ylabel={Ingestion time (seconds - Log Scale)}
        ]
    \addplot[mark=\RZSMark, mark size=4]
    table[x index=0,y index=1]{\testdata};
       \addlegendentry{RZS}
    \addplot[mark=\RDMark, mark size=4]
    table[x index=0,y index=2]{\testdata};
       \addlegendentry{Rasdaman}
    \addplot[mark=\GEEMark, mark size=4]
    table[x index=0,y index=3]{\testdata};
    \addlegendentry{Google Earth Engine}
    \end{loglogaxis}
\end{tikzpicture} \label{figure:ingestion_raster}}
    \subfigure[Vector Dataset]{\pgfplotstableread{ 
Vector      RZS	GEE
51638    4.169	120
165186	    2.302	60
3817412	2.677	600
38467094	    8.055	1320
52894188	    10.577	660
336145640	    70.193	7800
}\testdata

\begin{tikzpicture}[scale=0.45]
\begin{loglogaxis}
[legend style={at={(0.05,0.95)}, anchor=north west,legend columns=3},
    xlabel={No. of line segments},
    ylabel={Ingestion time (seconds - Log scale)}
    ]
\addplot[mark=\RZSMark, mark size=4]
table[x index=0,y index=1]{\testdata};
   \addlegendentry{RZS}
\addplot[mark=\GEEMark, mark size=4]
table[x index=0,y index=2]{\testdata};
\addlegendentry{Google Earth Engine}
\end{loglogaxis}
\end{tikzpicture} \label{fig:ingestion_vector}}
    \caption{Ingestion time}
    \label{figure:ingestion}
\end{figure}
\subsection{Ingestion Time}
Figure~\ref{figure:ingestion} shows the ingestion time of the raster and vector datasets for the proposed RZS algorithm, Rasdaman and GEE. Scanline method can read data from disk and hence does not have the overhead of ingestion time. RZS algorithm requires both raster and vector datasets to be ingested into HDFS, while GEE requires them to be uploaded to its web interface as well. We do not upload US Aster and Treecover to GEE as they are available in its data repository. Rasdaman only requires to ingest the raster datasets, although it failed to ingest the US Aster dataset as explained earlier. As can be observed from the Figure~\ref{figure:ingestion_raster} RZS has a lower raster data ingestion time as compared to Rasdaman and GEE. Figure~\ref{fig:ingestion_vector} shows that RZS has an order of magnitude lower vector data ingestion time when compared to GEE. The reason for that is that RZS follows an in-site data processing methodology which does not need to read the data until the query is executed. This makes it a perfect choice for ad-hoc exploratory queries.

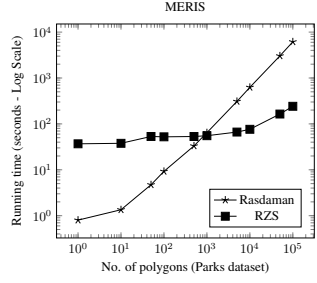
\begin{figure}
    \centering
    \pgfplotstableread{ 
polygons	Rasdaman	RZS
1	0.8007645607	36.83171688
10	1.347858429	37.77808137
50	4.720240116	53.21618234
100	9.247987032	52.27065339
500	33.15548468	53.13052013
1000	64.33489132	55.29434175
5000	308.9053059	66.41187233
10000	626.396395	76.31340747
50000	3034.97445	164.0000945
100000	6142.942665	241.9559255
}\testdata

\begin{tikzpicture}[scale=0.5]
\begin{loglogaxis}
[legend style={at={(0.95,0.05)}, anchor=south east},
title={MERIS},
xlabel={No. of polygons (Parks dataset)},
ylabel={Running time (seconds - Log Scale)}
]
\addplot[mark=\RDMark, mark size=3]
table[x index=0,y index=1]{\testdata};
   \addlegendentry{Rasdaman}
\addplot[mark=\RZSMark, mark size=3]
table[x index=0,y index=2]{\testdata};
   \addlegendentry{RZS}
\end{loglogaxis}
\end{tikzpicture}
    \caption{Scalability of Rasdaman and RZS}
    \label{fig:rasdaman_efficiency}
\end{figure}

\subsection{Closeup Scalability of Rasdaman}
Since Rasdaman preloads the raster data but not the vector data, it could be a good choice if only a few polygons need to be processed on a large raster dataset. Figure~\ref{fig:rasdaman_efficiency} shows the results of zonal statistics query using the MERIS dataset and varying the number of polygons in the parks dataset. It can be observed that Rasdaman is optimized for a very few polygons. However, its running time has a steep ascent as the number of polygons increase. This happens because Rasdaman processes each polygon in a separate query which results in overlapping work. RZS is able to scale more steadily as it can combine the overlapping work using the intersection file structure.

\subsection{Verification of Cost Models}
\label{sec:experiments:verification}

\begin{figure}[t]
    \centering
    \subfigure[GLC2000]{\pgfplotstableread{ 
VectorSize  DRZS    RZS
51638       6.63E+08	5.49E+01
165186      6.65E+08	4.96E+01
3817412     7.24E+08	5.67E+01
38467094    2.62E+09	6.39E+01
52894188    1.51E+09	5.88E+01
336145640   3.28E+10	9.05E+02
}\testdata

\begin{tikzpicture}[scale=0.45]
\begin{loglogaxis}
[title={Spearman's Correlation ($r_s=0.94286$)},
xlabel={Vector Size (no. of line segments)},
ylabel={Normalized cost (Log scale)}]
\addplot[mark=\DRZSMark, mark size=3]table[x index=0,y index=1]{\testdata};
\end{loglogaxis}

\begin{loglogaxis}
[legend style={at={(0.1,0.9)}, anchor=north west},
ticks=none]
\addplot[mark=\RZSMark, mark size=3]table[x index=0,y index=2]{\testdata};
\addlegendimage{mark=\DRZSMark}\addlegendentry{$D_{RZS}$ (Estimate)}
\addlegendimage{mark=\RZSMark}\addlegendentry{RZS (Actual)}
\end{loglogaxis}

\end{tikzpicture} \label{fig:verification:rzs-glc}}
    \subfigure[TreeCover]{\pgfplotstableread{ 
VectorSize  DRZS        RZS
51638       8.40E+11	9.60E+01	
165186      8.40E+11	7.53E+01	
3817412     8.40E+11	3.23E+02	
38467094    2.52E+12	2.93E+02	
52894188    8.42E+11	1.66E+02
336145640   3.11E+13	9.66E+03
}\testdata

\begin{tikzpicture}[scale=0.45]
\begin{loglogaxis}
[title={Spearman's Correlation $r_s=0.77$},
xlabel={Vector Size (no. of line segments)},
ylabel={Normalized cost (Log scale)}]
\addplot[mark=\DRZSMark, mark size=3] table[x index=0,y index=1]{\testdata};
\end{loglogaxis}

\begin{loglogaxis}
[legend style={at={(0.1,0.9)}, anchor=north west},
ticks=none]
\addplot[mark=\RZSMark, mark size=3]table[x index=0,y index=2]{\testdata};
\addlegendimage{mark=\DRZSMark}\addlegendentry{$D_{RZS}$ (Estimate)}
\addlegendimage{mark=\RZSMark}\addlegendentry{RZS (Actual)}
\end{loglogaxis}
\end{tikzpicture} \label{fig:verification:rzs-treecover}}
    \caption{Verification of the cost model of RZS}
    \label{fig:verification:rzs}
\end{figure}
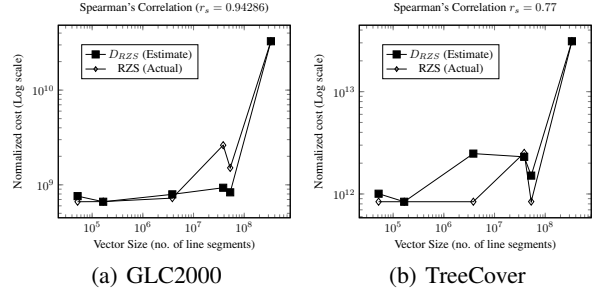

This part verifies our cost models described in Section~\ref{sec:fullanalysis} and uses that analysis to better explain the results that we got earlier. Since zonal statistics is a disk IO-intensive operation, we only use the disk cost estimation. Also, to be able to compare the actual running time to the estimated cost, we normalize all values and compare the trends rather than the absolute values. All cost models are computed based on the parameters in Table~\ref{tab:datasets}, the system parameters $B=128MB$ and $C=5,000$, and the equations in Section~\ref{sec:fullanalysis}.

Figure~\ref{fig:verification:rzs-glc} compares the actual running time of RZS to the estimated cost (both normalized) when processing GLC2000 and TreeCover datasets. As shown in figure, the trends generally match for both the small and the large datasets showing the robustness of the cost model. Notice that there is still some deviation due to our assumptions of uniform distribution of the vector data which does not hold in reality. To quantify the relationship, we calculated Spearman's correlation coefficient for both cases and it turned out to be $0.94$ and $0.77$ for GLC2000 and TreeCover respectively. Notice that we did not use the Pearson's correlation coefficient as it will result in a {\em misleading} value of almost 0.999 due to the exponential increase on the the $x$ and $y$ values.

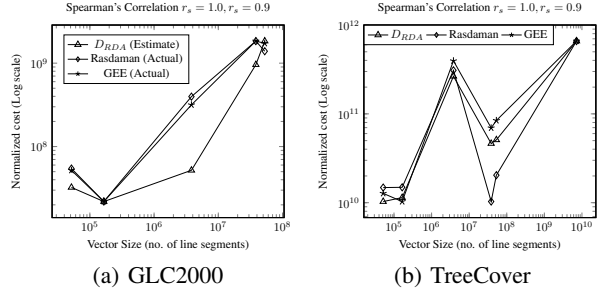
\begin{figure}[t]
    \centering
    \subfigure[GLC2000]{\pgfplotstableread{ 
VectorSize  DRDA	Rasdaman GEE
51638       5.45E+07	1.45E+02	2.20E+01	
165186      2.19E+07	5.61E+01	1.67E+01	
3817412     3.97E+08	1.06E+03	3.07E+01	
38467094    1.84E+09	7.36E+03	2.40E+02	
52894188    1.39E+09	6.87E+03	3.80E+02
}\testdata	

\begin{tikzpicture}[scale=0.45]
\begin{loglogaxis}
[title={Spearman's Correlation $r_s=1.0, r_s=0.9$},
xlabel={Vector Size (no. of line segments)},
ylabel={Normalized cost (Log scale)}]
\addplot[mark=\DRDAMark, mark size=3] table[x index=0,y index=1]{\testdata};
\end{loglogaxis}

\begin{loglogaxis}
[ticks=none]
\addplot[mark=\RDMark, mark size=3] table[x index=0,y index=2]{\testdata};
\end{loglogaxis}

\begin{loglogaxis}
[ticks=none,
legend style={at={(0.05,0.95)}, anchor=north west},
]
\addplot[mark=\GEEMark, mark size=3] table[x index=0,y index=3]{\testdata};
\addlegendimage{mark=\DRDAMark}\addlegendentry{$D_{RDA}$ (Estimate)}
\addlegendimage{mark=\RDMark}\addlegendentry{Rasdaman (Actual)}
\addlegendimage{mark=\GEEMark}\addlegendentry{GEE (Actual)}
\end{loglogaxis}

\end{tikzpicture} \label{fig:verification:rda-glc}}
    \subfigure[TreeCover]{\pgfplotstableread{ 
VectorSize  DRDA        Rasdaman    GEE
51638       1.48E+10	9.26E+02	1.14E+02	
165186      1.49E+10	6.76E+02	1.30E+02	
3817412     3.09E+11	1.52E+05	7.23E+03	
38467094    1.03E+10	1.15E+04	7.80E+02	
52894188    2.04E+10	1.53E+04	8.80E+02
7172610906  6.60E+11	3.25E+05	2.32E+04
}\testdata

\begin{tikzpicture}[scale=0.45]
\begin{loglogaxis}
[title={Spearman's Correlation $r_s=1.0, r_s=0.9$},
xlabel={Vector Size (no. of line segments)},
ylabel={Normalized cost (Log scale)}]
\addplot[mark=\DRDAMark, mark size=3] table[x index=0,y index=1]{\testdata};
\end{loglogaxis}

\begin{loglogaxis}
[ticks=none]
\addplot[mark=\RDMark, mark size=3] table[x index=0,y index=2]{\testdata};
\end{loglogaxis}

\begin{loglogaxis}
[ticks=none,
legend style={at={(0.00,1.00)}, anchor=north west,legend columns=3},
]
\addplot[mark=\GEEMark, mark size=3] table[x index=0,y index=3]{\testdata};
\addlegendimage{mark=\DRDAMark}\addlegendentry{$D_{RDA}$}
\addlegendimage{mark=\RDMark}\addlegendentry{Rasdaman}
\addlegendimage{mark=\GEEMark}\addlegendentry{GEE}
\end{loglogaxis}

\end{tikzpicture} \label{fig:verification:rda-treecover}}
    \caption{Verification of the cost model of RDA}
    \label{fig:verification:rda}
\end{figure}

Figure~\ref{fig:verification:rda} shows the estimated cost of RDA and the actual cost of both Rasdaman and GEE. It is evident from the chart that the trend of the the cost model and the actual times are very similar. While we do not have definite information about GEE, we highly predict that they use the RDA algorithm given the almost perfect match in trends. The correlation coefficient with Rasdaman in GLC2000 is perfect, and for GEE 0.9 and 0.83. We did not have enough data points for Rasdaman with the TreeCover dataset to plot the figure or calculate the coefficient. We believe the cost model for RDA is more accurate since it does not make an assumption about the uniformity of the data as we do with RZS. Still, it is amazing how accurate the results are given that we only relied on the statistics shown in the Table~\ref{tab:datasets}.

\subsection{Applications}
\label{sec:experiments:applications}
This section discusses two real-life applications of our proposed system, {\em population estimation} and {\em wildfire combating}.


The first application of RZS is to estimate the population of arbitrary regions using landcover data~\cite{reibel2007areal}. The problem is that the US Census Bureau reports the population at the granularity of {\em census tracts} which are regions chosen by the Bureau to keep the privacy of the data. Areal interpolation transforms these counts from source polygons, i.e., tracts, to target polygons, e.g., ZIP Codes, with unknown counts. One accurate method~\cite{reibel2007areal} uses the National Land Cover database (NLCD)~\cite{nlcd} raster dataset as a reference to disaggregate the population counts into pixels and then aggregate them back into target polygons. To speed up the process, we apply RZS to compute the histogram for each polygon in the TRACT dataset on the NLCD dataset. We compared our single-machine implementation to the original Python implementation used by the developers of that algorithm, which was a vector database approach (VDA). Using RZS, the entire process completed in 10 seconds for the state of Pennsylvania while the python-based script took over 100 minutes to complete. Given that impressive speedup, the authors were able to scale their work to the entire US which took under 2 hours on a single machine. 

The second application of RZS is in combating wildfires. The goal of this application is to co-create probabilistic decision theoretic models to combat wildfires, which would use RZS for pre-processing satellite wildfire data. The raster data used has a size of 60~GB and the vector data has over 3 million polygons, both spanning over California. We have been able to compute zonal statistics for it in under 2 hours using a AWS cluster of 20 machines.

\subsection{Tuning}
\label{sec:experiments:tuning}

This section provides a deeper study of the proposed algorithm and examines the effectiveness of each component individually.

\begin{figure*}[t]
\centering
\includegraphics[width=\textwidth]{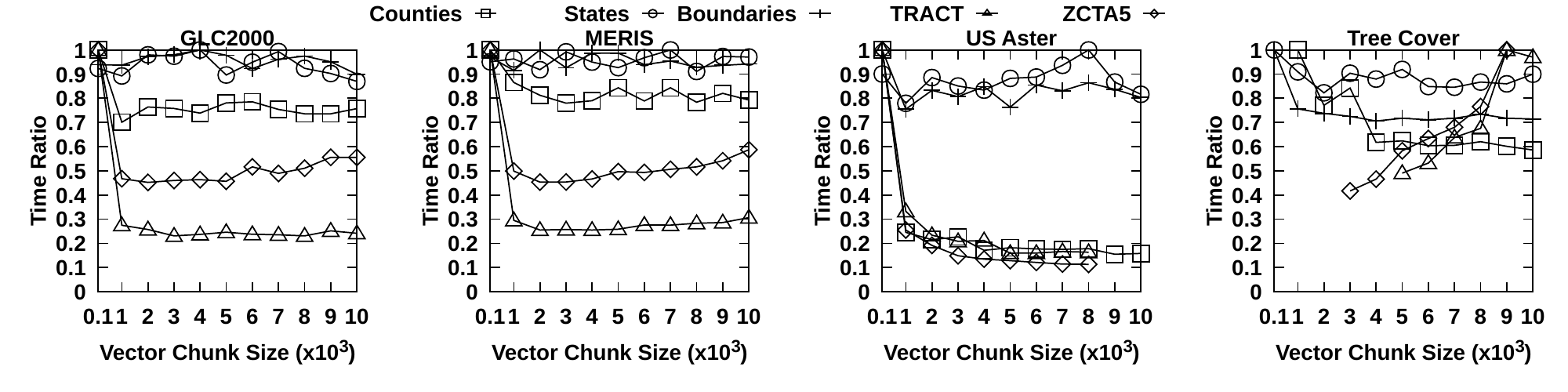}
\caption{Effect of different sizes of Vector Chunks on total running time}
\label{figure:vectorchunks}
\end{figure*}

\subsubsection{Vector Chunks}
\label{sec:experiments:tuning:vector chunks}
This experiment studies the effect of splitting the vector file into chunks.
In particular, this experiment varies the sizes of vector chunk used in RZS starting with 100, then 1000 to 10,000, incremented in steps of 1000. Figure~\ref{figure:vectorchunks} shows the overall running time as the chunk size increases. Since each line in this graph represents a different raster dataset, we are only interested in the trend of the lines. Therefore, each line is normalized independently to fit all of them in one figure. We omit the running times for which the algorithm runs out of memory. We observe in this experiment that a very low chunk size of 100 results in a reduction in the performance due to the overhead of creating and running too many RaptorSplits. On the other hand, using a very large chunk size eventually results in some job failures due to the memory overhead. This is equivalent to not splitting the vector file.

After chunk size of 3000, the number of vector chunks generated for Counties and Boundaries, become stable which leads to marginal variation in their running times. The variation of chunk size on larger vector datasets and Tree Cover is more prominent than the other vector datasets. This is due the large size of the Tree Cover dataset. The increase in vector chunk size leads to a decrease in the number of chunks being generated and hence, less number of \emph{raptor splits}. This leads to each machine having more amount of work to do, and a non-optimal distribution of work.
It can be concluded that the choice of vector chunk size should neither be too big (10,000) or too small(100). It should lead to an optimal distribution of work in the \emph{Step 2: Raptor Data Processing} and can depend on the system configuration. We chose it to be 5,000 based on the experiments and according to our system configuration. 

\subsubsection{Compression of Intersection File}
\label{sec:experiments:tuning:compression}

\begin{table}[t]
\centering
\caption{Compression Ratio of Intersection Files}
\centering
\small
\begin{tabular}{|c|c|c|c|c|}
\hline
& GLC2000 & MERIS & US Aster & Tree Cover\\
\hline
Counties & 4.13 & 4.18 & 4.34 & 5.22\\
\hline
States  & 3.41 & 3.34 & 3.3 & 6.07\\
\hline
Boundaries & 2.56 & 2.55 & 3.68 & 17.43\\
\hline
TRACT   & 2.94 & 3.21 & 3.45 & 4.9\\
\hline
ZCTA5 & 2.88 & 3.11 & 3.34 & 4.64\\
\hline
\end{tabular}

\label{tab:intersections}
\end{table}

\begin{figure}
		\begin{center}
		\includegraphics[width=0.5\columnwidth]{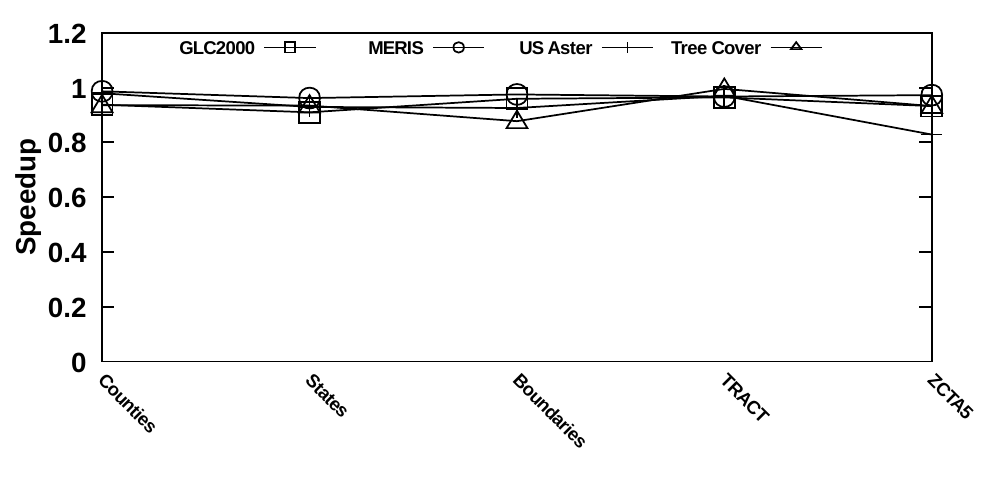}
		\caption{Effect of compression of Intersection File}
		\label{figure:compression}
		\end{center}
\end{figure}

In this experiment, we study the effect and trade-off of compressing the intersection files. The size of the raw (uncompressed) intersection files ranges from a few kilobytes to a gigabyte. In order to reduce the network and disk IO when storing this file in the distributed file system, we investigated the option of compressing the intersection files using both GZIP and Snappy compression libraries. We did not see a major difference between the two libraries so we are only reporting the results of GZIP. The speedup of the approach without using compression to the approach with the use of compression can be seen in Figure~\ref{figure:compression}. It can be observed that the speedup is either equal to one or marginally less than it. This means that the total running time without the use of compression is less than or almost equal to that with compression.
Table~\ref{tab:intersections} reports the compression ratio of intersection files, which is defined as the ratio of the size of uncompressed file to that of the compressed file. Although, from Table~\ref{tab:intersections}, it can be seen that the size of compressed files is far smaller than that of non-compressed \emph{intersection files} this did not provide a significant improvement in the running time. This is because the time saved in writing the compressed intersection file is nearly the same as the time taken for compression and decompression of the intersection file. However, compressing the intersection file can be a viable option, in case network IO becomes a bottleneck.

\subsubsection{Spatial Partitioning of Vector Data}

\begin{figure}[t]
\begin{center}
\includegraphics[width=0.7\columnwidth]{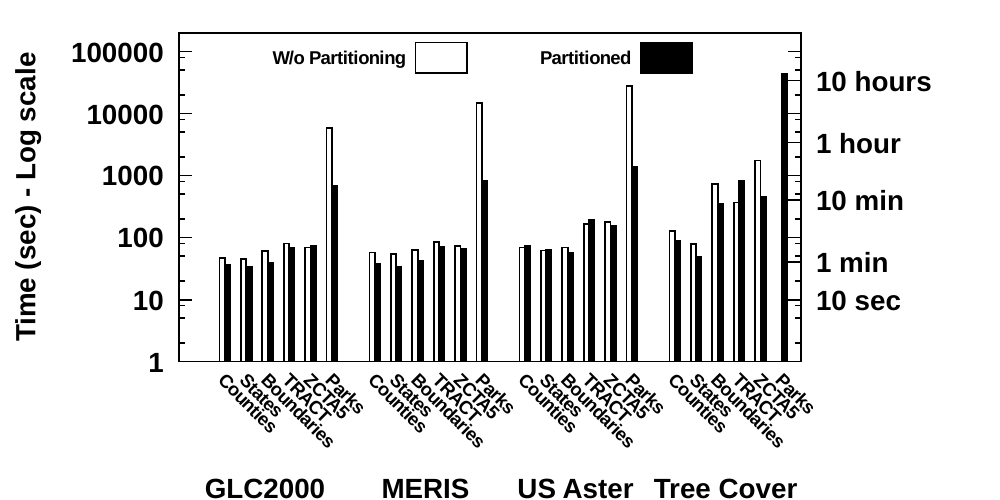}
\caption{Effect of spatially partitioning vector data}
\label{figure:indexing}
\end{center}
\end{figure}

This experiment shows the effect of spatial partitioning input vector data on the total running time. We use R*-Grove~\cite{VE18} algorithm to partition the vector data. The results for the total running time of the proposed algorithm with partitioned and non-partitioned vector data are shown in figure~\ref{figure:indexing}. As can be observed from the figure, spatially partitioning the vector data helps achieve orders of magnitude of speedup for the large vector datasets (ZCTA5 and Parks). In addition, the largest combination of datasets, Parks$\times$TreeCover took more than 48 hours with non-partitioned data and we had to terminate it. The total running time for other vector datasets either shows a marginal improvement or remains the same.

Spatial partitioning reduces the spatial extents of the contents of each HDFS block which also reduces the overlap with raster data. Not using spatial partitioning means that the contents of each block would cover the entire input space which results in reading the entire raster file when processing each intersection file. Our cost model can further explain the results of this experiment by setting the average width and height of a block to the width and height of the raster dataset, respectively. For small vector datasets with one block, there will be no difference between spatial and non-spatial partitioning. However, as the number of blocks increase the effect of spatial partitioning will be huge.


\begin{table}
\begin{center}
\caption{Time taken to Partition Vector Datasets}
\flushleft
\begin{center}
\normalsize
\begin{tabular}{|c|c|}
\hline
Vector Dataset & Partitioning Time\\
\hline
Counties & 22.97s\\
\hline
States  & 38.92s\\
\hline
Boundaries & 121.09s\\
\hline
TRACT   & 118.14s\\
\hline
ZCTA5 & 126.85s\\
\hline
Parks & 209.40s\\
\hline
\end{tabular}
\end{center}
\label{tab:indexing}
\end{center}
\end{table}

Table~\ref{tab:indexing} reports the time taken by R*-Grove~\cite{VE18} algorithm to partition the vector datasets. As can be seen from the table, 
The partitioning though fast takes hundreds of seconds to partition the vector dataset. Parks was the only dataset, where the difference between running time of partitioned and non-partitioned input vector data is much larger than the partitioning time. It can thus be concluded that partitioning vector data is of advantage for very large vector datasets.

\section{Conclusion}
\label{sec:conclusion}

This paper models the zonal statistics problem as a join problem and propose a distributed MapReduce algorithm, Raptor Zonal Statistics (RZS) to solve it. The proposed algorithm provides several key ideas that can carry on to other distributed algorithms for processing the combination of big vector and raster datasets. First, the proposed framework runs in three steps, an intersection step that computes a common distributed index structure, a selection step that reads the pixels from the raster file based on that structure, and an aggregation step that processes the pixel values. The second key idea is introducing the RaptorInputFormat which is the first input format that combines raster plus vector data in one split. The RaptorInputFormat can define the units of work to be executed in parallel and provides an easy way to optimize and balance the load across machines. The RaptorInputFormat combined with the intersection step can also prune irrelevant parts of the raster file in order to speed up the parallel processing. Our experiments with large scale real data shows that the proposed algorithm is up-to two orders of magnitude faster than the baselines including Rasdaman and Google Earth Engine (GEE). We also presented a cost model which helped us explaining the results of both RZS and the baseline techniques.

\bibliographystyle{unsrt}
\bibliography{references.bib}
\appendix

\end{document}